\documentclass[a4paper]{jpconf}
\usepackage{graphicx}
\usepackage{amsmath,amsfonts,amsthm,amssymb} 
\usepackage[makeroom]{cancel}
\begin{document}
\title{Expanding Core-Collapse Supernova Search Horizon of Neutrino Detectors}

\author{Odysse Halim$^{a,b,*}$, C Vigorito$^{c,d}$, C Casentini$^{e,f}$, G Pagliaroli$^{a,b}$, M Drago$^{a,b}$, V Fafone$^{f,g}$}

\address{$^a$Gran Sasso Science Institute (GSSI), L'Aquila, Italy\\ $^b$INFN sezione di Laboratori Nazionali del Gran Sasso (LNGS), Assergi, Italy\\ $^c$University of Turin, Turin, Italy\\ $^d$INFN (Istituto Nazionale Fisica Nucleare) sezione di Torino, Turin, Italy\\ $^e$INAF - IAPS, Rome, Italy\\ $^f$INFN (Istituto Nazionale Fisica Nucleare) sezione di Roma Tor Vergata, Rome, Italy\\ $^g$University of Rome Tor Vergata, Rome, Italy\\ \textsuperscript{*}\textit{corresponding author}}




\begin{abstract}
Core-Collapse Supernovae, failed supernovae and quark novae are expected to release an energy of few $10^{53}$ ergs through MeV neutrinos and a network of detectors is operative to look online for these events. However, when the source distance increases and/or the average energy of emitted neutrinos decreases, the signal statistics drops and the identification of these low statistic astrophysical bursts could be challenging. In a standard search, neutrino detectors characterise the observed clusters of events with a parameter called multiplicity, i.e. the number of collected events in a fixed time-window. We discuss a new parameter called $\xi$ (=multiplicity/duration of the cluster) in order to add the information on the temporal behaviour of the expected signal with respect to background. By adding this parameter to the multiplicity we optimise the search of astrophysical bursts and we increase their detection horizon. Moreover, the use of the $\xi$ can be easily implemented in an online system and can apply also to a network of detectors like SNEWS. For these reasons this work is relevant in the multi-messengers era when fast alerts with high significance are mandatory.  
\end{abstract}


\section{Introduction}
Several astrophysical sources are expected to give rise to impulsive bursts of low-energy neutrinos \cite{casentini} such as core-collapse supernovae (CCSNe) \cite{janka}, failed supernovae (failed-SNe) \cite{adams}, as well as the hypothetical quark novae (QNe) \cite{ouyed}. All  these sources have a common signature, which is an impulsive neutrino emission with a total emitted energy of the order of $10^{53}$ ergs,  average neutrino energy of $\sim 10$ MeV and a short duration of the order of 10 seconds ($\sim0.5$ seconds for the failed-SNe).

The impulsive neutrino emission can be detected by several running neutrino detectors. In this work we will focus on LVD, KamLAND, and Super Kamiokande detector and we will exploit Monte Carlo background simulations based on the features reported in table \ref{tab:detector}. 
\begin{center}
\begin{table}[h]
\centering
\caption{\label{tab:detector}Specifications of the involved detectors.} 
\begin{tabular}{@{}l*{15}l*{15}l*{15}{l}}
\br
Detectors& LVD& KamLAND&SuperK\\
\mr
Type& Liquid Scin. & Liquid Scin. & Water Cherenkov \\
\mr
Channel, mass & IBD, $\bar{\nu}_e$, 1 kton &IBD, $\bar{\nu}_e$, 1 kton & IBD, $\bar{\nu}_e$, 22.5 kton\\
\mr
$f_\mathrm{bkg}$, $E_\mathrm{thr}$ &0.028 Hz, 10 MeV& 0.015 Hz, 1MeV& 0.012 Hz, 7 MeV\\
\br
\end{tabular}
\end{table}
\end{center}



\section{The modified imitation frequency}
In low energy neutrino analysis to search for CCSNe, a temporal data set from a detector is binned in a sliding time window of $w=20$ seconds. 
The group of events inside each window is defined as a \textit{cluster} and the number of events in the cluster is called multiplicity $m_i$. The multiplicity distribution due to background only events is supposed to follow Poisson distribution and the $i$-th cluster \textit{imitation frequency}\footnote{This imitation frequency can be also understood as \textit{false-alarm-rate}, FAR.} ($f^{im}$), correlated with the significance, is defined as:
\begin{equation}
f^{im}_i= 8640 \times \sum_{k=m_i}^\infty \frac{(f_{bkg}\cdot w)^k e^{-f_{bkg}\cdot w}}{k\!} \mathrm{[day^{-1}]}=8640\times \sum_{k=m_i}^\infty P(k)
\label{eq:old_fim}
\end{equation}
where 8640 is the total number of bins in one day being the search performed, at first, in consecutive 20 second bins, and then by shifting the temporal bins of 10 seconds in order to eliminate boundary problems. In this formula $P(k)$ represents the probability that a cluster of multiplicity $k$ is produced by the background.

Based on \cite{casentini}, we also characterised each cluster by its duration, i.e. the time elapsing from the first to the last event in the cluster. This duration of a cluster, thus, can range from 0 to 20 seconds. We introduce the new-added parameter ($\xi$) for each cluster defined as $\xi\equiv \frac{[\mathrm{multiplicity}]}{[\mathrm{duration}]}$. The results of \cite{casentini} showed that by performing a new additional cut in $\xi$, it is possible to disentangle further the simulated astrophysical signals from background.

\begin{figure}[!ht]
\centering
\includegraphics[width=.5\textwidth]{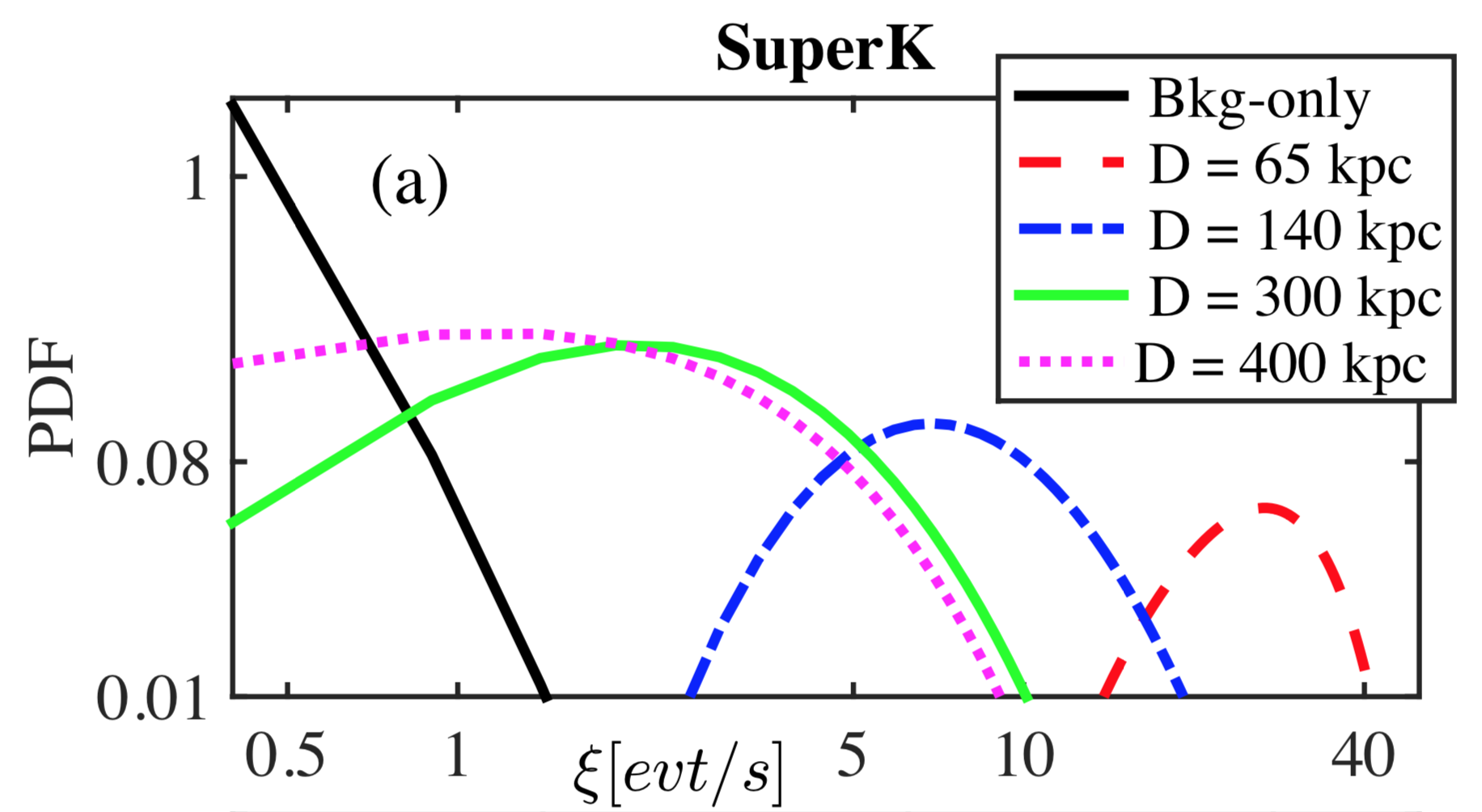}
\caption{Probability density functions for background plus signal clusters as functions of the parameter and for different distances in the case of SuperK detector. The black solid line shows the PDF of pure background clusters. Figure is taken from \cite{casentini}.}
\label{fig:pdf}
\end{figure}

In this work, instead of using $\xi$ as a hard cut, we can define a two parameter ($m_i$ and $\xi$) imitation frequency for each cluster, called $F^{im}$ (capital letter), calculated as the following: 
\begin{equation}
F^{im}_i= 8640 \times \sum_{k=m_i}^\infty P(k) \int_{\xi=\xi_i}^\infty P(\xi\geq\xi_i|k) d\xi.
\label{eq:newfim_1}
\end{equation}
In this equation the numerical factor is the same as in equation \ref{eq:old_fim}, while the term inside the sum represents the joint probability $P(k,\xi)$, i.e. the probability that a cluster with multiplicity $k$ \textbf{and} a specific value for $\xi$ is produced by the background. This joint probability can be reduced to $P(k,\xi)=P(\xi | k)P(k)$, thanks to the total probability law where $P(\xi | k)$ is the conditional probability that a cluster with multiplicity $k$ is present with a specific $\xi$ value. 
The conditional probability can be derived for each detector by taking into account the distribution of the $\xi$ values from clusters only due to background, following \cite{casentini}. As a leading example the distribution for Super Kamiokande is reported in figure \ref{fig:pdf} in the form of a probability density functions (PDF).
The black solid line in the figure represents the normalised probability for background clusters to have a specific value of $\xi$ so that,
\begin{equation}
\int_{k/20}^\infty N_kf(\xi)d\xi=\int_{k/20}^\infty P( \xi\geq\xi_i | k)d\xi=1,
\label{eq:pdf_func}
\end{equation}
where $k$ in this case is a \textit{possible} multiplicity value. In \cite{casentini}, it has been observed that the background curve has a probability density function (PDF) of 4-parameter Gamma distribution with some $p$-value of the fit. However, in order to be more generic, we will not use any distribution fit and we rely on the \textbf{interpolation} of the simulated background.


We argue that simulated injections should have smaller values of imitation frequency when we use the new method while background cluster imitation frequencies should be unchanged. We know for sure that the cluster multiplicity is never 0 inside this 20-second bin and $\xi_\mathrm{min}$ has a value of $k/\mathrm{[max\,\,duration]}\rightarrow k/20$. Then, we can rearrange equation \ref{eq:pdf_func} with the normalisation factor written as $N_k=1/\int_{k/20}^\infty f(\xi)d\xi$. Therefore, the conditional probability becomes,
\begin{equation}
P(\xi | k) = P(\xi\geq\xi_i | k) =  1-\frac{\int_{k/20}^{\xi\geq k/20}   f(\xi) d\xi}{\int_{k/20}^\infty f(\xi)d\xi},
\end{equation}
Thus the new imitation frequency can be written as,
\begin{equation}
\boxed{F_i^{im}=  \left[f_i^{im}-8640\times\sum\limits_{k=m_i}^{m_i+n;\,n\leq (20\xi-m_i)} P(k)N_{k}\int_{k/20}^\xi f(\xi)d\xi \right]\mathrm{[day^{-1}]}.}
\label{eq:newfim}
\end{equation}
All in all, this new imitation frequency will work as long as we have a definite background distribution of $f(\xi)$ for each detector.

\section{Results}
\begin{figure}
\centering
  \includegraphics[width=.6\linewidth]{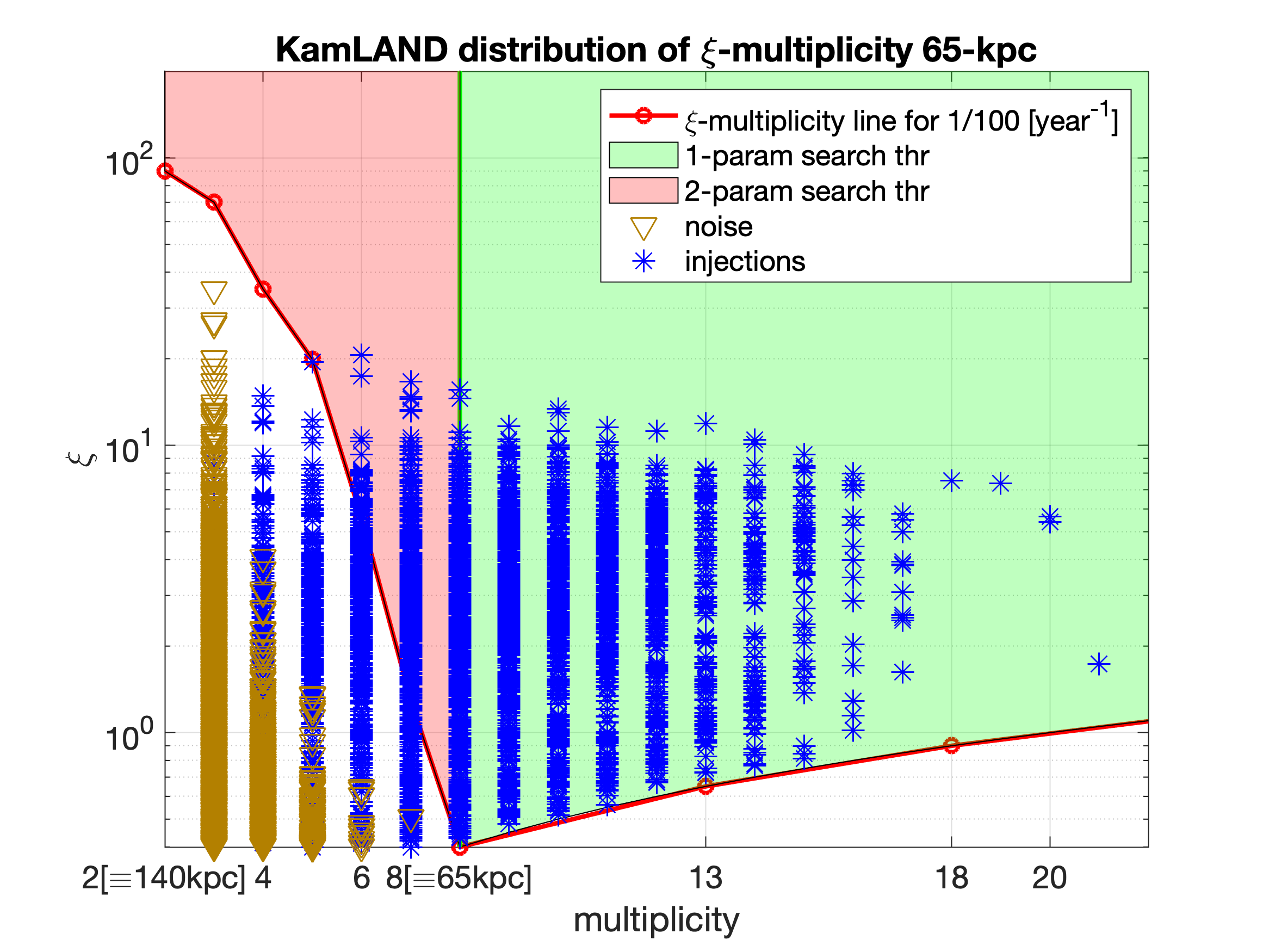}
  \caption{KamLAND $\xi$-multiplicity map with 10-year simulated background and 65-kpc injections.}
  \label{fig:kam_bkg_inj}
\end{figure}

We reanalysed simulated KamLAND background + injections of 65 kpc from \cite{casentini} and we obtained more recovered injections (without adding any more noise) with our 2-parameter method. The comparison between the two methods is in figure \ref{fig:kam_bkg_inj}: the green area is the standard 1-parameter search ($m$-only) while the red area is the improvement for our 2-parameter method ($m,\xi$) where both areas pass the threshold in imitation frequency\footnote{equation \ref{eq:old_fim} is the imitation frequency for the 1-parameter while equation \ref{eq:newfim} for the 2-parameter.} of $\leq1/100$ years following the SNEWS \cite{snews} criteria. Several recovered injections lie on the red area, meaning that they are recovered by the new method but missed in the standard method. More quantitative information can be seen in table \ref{tab:kam_single} for the efficiency ($\equiv[\mathrm{recovered}]/[\mathrm{total\,injections}]$) of both methods (column 3 and 4), while no noise surpassing the threshold (column 2). 
\begin{table}
\caption{Single detector KamLAND 65-kpc}
\label{tab:kam_single}
\centering
{\renewcommand{\arraystretch}{1.2}
\begin{tabular}{||c | c | c | c ||} 
\hline
Noise & Noise   & Efficiency 1-parameter & Efficiency  2-parameter \\
 & $\left[1/100\,\mathrm{years}\right]$ &  $\left[<1/100\,\mathrm{years}\right]$ & $\left[<1/100\,\mathrm{years}\right]$ (Our Method)  \\
\hline
75198 & 0\% = 0/75198 & \textbf{59.0\%} = 2155/3654 & \textbf{70.6\%} = 2581/3654   \\
\hline
\end{tabular}
}
\end{table}

In addition, we also have applied the coincidence analysis following \cite{casentini} with the simulations of KamLAND-LVD and the result is in table \ref{tab:kam_lvd_det}. In this case, we compare the $5\sigma$ efficiency. This $5\sigma$ is instead the threshold for the {\it false-alarm-probability} $\mathrm{FAP}=1-\exp{(-\mathrm{FAR}\times\mathrm{livetime})}$ with the FAR is defined in c.f. equation 4.1 in \cite{casentini} and the livetime is the common observing time between the data of KamLAND and of LVD.
\begin{table}
\caption{Efficiency and misidentification probability for KamLAND-LVD.}
\label{tab:kam_lvd_det}
\centering
{\renewcommand{\arraystretch}{1.2}
\begin{tabular}{| c || c | c  |}
 \hline
 2-detector: & \multicolumn{2}{c|}{10 year - 65 kpc} \\
  \cline{2-3}
LVD - KamLAND& 1-parameter & 2-parameter  \\
\hline
$5\sigma$ Efficiency & $\mathbf{62.9\%} = 2298/3654$ & $\mathbf{80.8\%} = 2951/3654$ \\
\hline
$5\sigma$ Misidentification Prob & $0\% = 0/3872$ & $0\% = 0/3872$ \\
 \hline
\end{tabular}
}
\end{table}

Based on our result, the new 2-parameter method could improve the efficiency of current neutrino detectors for CCSN search. Moreover, this may also improve our capability for failed-SN search till Large/Small Magellanic Cloud by SuperK together with GW data from LIGO-Virgo.



\section*{References}
\begin{thebibliography}{9}
\bibitem{casentini} Casentini C et.al. 2018 JCAP \textbf{1808} (2018) 010. DOI:10.1088/1475-7516/2018/08/010.
\bibitem{janka} Janka H -T et.al. 2007 Phy. Rep. \textbf{442} 1-6, April 2007, p.38-74. DOI:10.1016/j.physrep.2007.02.002.
\bibitem{adams} Adams S M et.al. 2017 Mon. Not. Roy. Astron. Soc. \textbf{468} (2017) 4, 4968-4981. DOI:10.1093/mnras/stx816.
\bibitem{ouyed} Ouyed R et.al. 2002 Astron. Astrophys. \textbf{390} (2002) L39. DOI:10.1051/0004-6361:20020982.
\bibitem{snews} Antonioli P et.al. 2004 New J. Phys. \textbf{6} (2004) 114. DOI: 10.1088/1367-2630/6/1/114.
\end{thebibliography}
\end{document}